%% file: main.tex
\documentclass[letterpaper,twocolumn,10pt]{article}

\input{preamble}

\usepackage{usenix2019_v3}
\usepackage{tikz}
\usepackage{amsmath}
\usepackage{comment}

\usepackage{filecontents}

\usepackage{kotex}
\usepackage[normalem]{ulem}
\usepackage{xcolor}

\begin{document}

\date{}

\title{\Large \bf VS-Flash: Harnessing Vector Search with Computational Memory-Semantic Flash}

\author{
  {\rm Donguk Kim}\thanks{Corresponding author: Donguk Kim (donguk12.kim@samsung.com).}\\
  Samsung Semiconductor
  \and
  {\rm Sangjun Pae}\\
  Samsung Semiconductor
  \and
  {\rm Daeyoung Lee}\\
  Samsung Semiconductor
  \and
  {\rm Dong Kim}\\
  Samsung Semiconductor
  } 

\maketitle

\begin{abstract}

As generative AI continues to evolve, retrieval-augmented generation (RAG) systems are driving rapid growth in vector database capacity requirements. As datasets outgrow host DRAM, vector databases rely on MMap-based serving over SSDs to extend capacity. However, high-fan-out ANN searches generate concurrent irregular accesses that contend for host memory and storage resources, degrading bulk throughput and inflating tail latency under mixed workloads.

In this paper, we address these challenges through memory-semantic capacity and workload-aware compute placement. \name exposes flash as a large-capacity memory-semantic space and dynamically selects the execution engine according to runtime query and workload characteristics. It offloads throughput-oriented large-batch queries to a Computational Memory-Semantic Flash (CMS-Flash) device, while executing latency-sensitive small-batch queries on the host CPU. Both paths access the same resident segments, avoiding a separate host-DRAM replica.

Across three billion-scale datasets, \name improves bulk-search throughput over vanilla Milvus by 36.8$\times$ on average, with up to a 57.4$\times$ improvement. Under mixed workloads, \name reduces the p99 latency of latency-sensitive requests by up to 24.8$\times$ while sustaining device-side bulk search. These results show that \name delivers scalable vector search beyond host DRAM capacity while jointly improving bulk throughput and small-request tail latency.

\end{abstract}

\input{contents/1-intro.tex}
\input{contents/2-background.tex}

\input{contents/3-design.tex}
\input{contents/5-evaluation.tex}

\input{contents/7-relatedwork.tex}

\input{contents/6-conclusion}

\bibliographystyle{plain}
\bibliography{refs.bib}

\end{document}

%% file: preamble.tex
\usepackage{kotex}
\usepackage{bm}
\usepackage{blindtext}
\usepackage{booktabs}
\usepackage{listings}
\usepackage{xcolor}
\usepackage{bold-extra}
\usepackage[export]{adjustbox}
\usepackage{algorithm}
\usepackage{algcompatible}
\usepackage{xspace}
\usepackage{amsmath}
\usepackage{tabu}
\usepackage{xcolor}
\usepackage[htt]{hyphenat}

\newcommand{\SmartIncludeGraphics}[3]{%
    \IfFileExists{#1}{\includegraphics[width=#2,height=#3,keepaspectratio]{#1}}{\fbox{\rule{0pt}{#3}\rule{#2}{0pt}}}%
  }

\usepackage{adjustbox}
\usepackage{tikz}
\usepackage[scaled=0.80]{beramono}
\usepackage[T1]{fontenc}

\usepackage{multirow}

\usepackage[nocompress]{cite}

\newcommand{\name}{VS-Flash\xspace}
\newcommand{\cmsflash}{CMS-Flash\xspace}

\newcommand{\memflash}{{Memory-Semantic Flash}\xspace}
\newcommand{\segmentmanager}{{Preload Manager}\xspace}
\newcommand{\router}{{Batch Dispatcher}\xspace}
\newcommand{\cxl}{CXL\xspace}

\newcommand{\mmapio}{{MMap-enabled Data Storage}\xspace}

\newcommand{\para}[1]{\noindent \textbf{#1.}}

%% file: contents/1-intro.tex
\section{Introduction}

Vector similarity search is a core operation in retrieval-augmented generation (RAG), recommendation, and semantic retrieval~\cite{athena, surveyofVDBMS}. Vector database management systems such as Milvus~\cite{wang2021milvus} commonly serve approximate nearest-neighbor (ANN) indexes through the host memory hierarchy~\cite{wang2021milvus, guo2022manu, surveyTech}. When a vector-search dataset exceeds the DRAM capacity available to a query node, Milvus's \mmapio feature retrieves index pages and vectors on demand. The resulting fine-grained accesses during irregular ANN traversal contend for limited DRAM capacity and repeatedly invoke host-OS page placement and page-fault handling.

\name replaces demand paging with a segment-centric, data-centric execution model in which segment residency governs both data movement and computation placement. It explicitly stages whole active segments from flash in device DRAM, then executes large-batch searches on device-side many-core resources beside the resident data. The same segments remain directly accessible to the host CPU, allowing latency-sensitive small batches to execute without a separate host-DRAM replica. This organization couples computation offload with shared segment access, rather than treating data placement and execution as independent decisions.

In summary, we make the following contributions:

\begin{itemize}
\item We identify asymmetric interference in Milvus MMap serving: throughput-oriented large batches intensify host-side memory contention and degrade the latency of small batches under mixed workloads.

\item We design \name, which preloads whole segments into device DRAM, executes bulk search on a CMS-Flash device, and routes latency-sensitive requests to the host CPU without a host-DRAM replica.

\item We implement \name in Milvus with a device-side ANN runtime and evaluate bulk-search throughput and mixed-workload tail latency.
\end{itemize}

%% file: contents/2-background.tex
\section{Background and Motivation}

\subsection{Milvus Query Processing}

\begin{figure}[t]
\centering
\includegraphics[width=0.95\columnwidth]{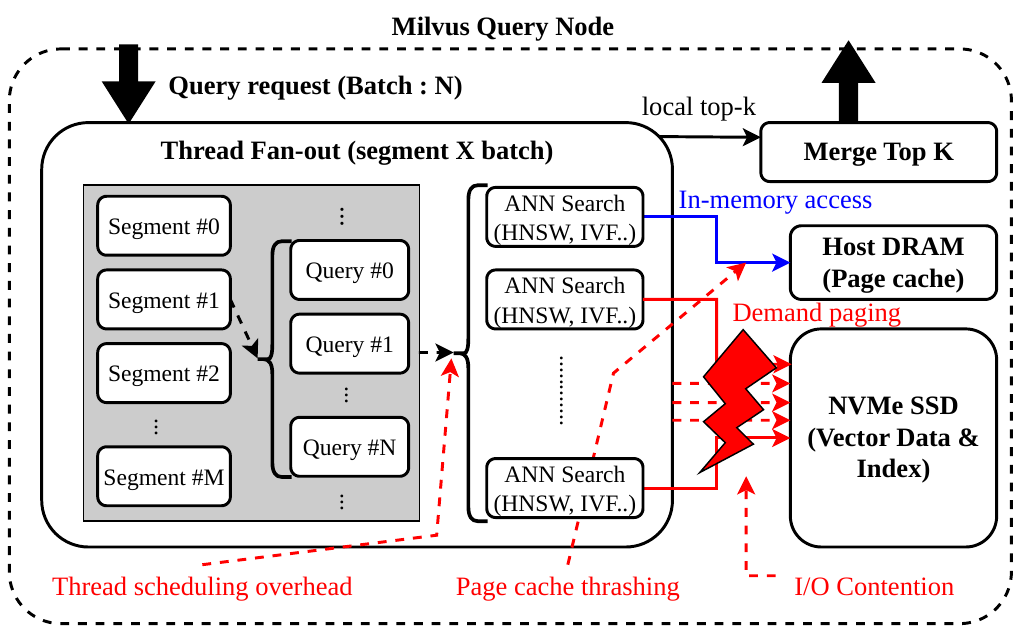}
\caption{Illustration of the Milvus query process with \mmapio.}
\label{fig:milvus_overview}
\end{figure}

\para{Segment-level Query Fan-out} Milvus partitions a collection into segments, each maintaining an independent vector index and serving as a unit of data organization and query execution. Figure~\ref{fig:milvus_overview} illustrates how a Milvus Query Node processes a query batch. The Query Node splits a batch into sub-searches for the target segments. Each sub-search is assigned to an independent thread and produces a local top-$k$ candidate list. The Query Node then merges the per-segment candidates into a final top-k result~\cite{wang2021milvus, guo2022manu}. 

For an HNSW index, each sub-search follows query-dependent graph neighbors within its target segment and repeatedly accesses index and vector data along an irregular traversal path~\cite{hnsw}. As the number of target segments or queries in a batch grows, the Query Node must create and coordinate more independent search tasks. This fan-out increases thread scheduling overhead before the search work itself completes.

\para{\mmapio} Milvus conventionally keeps the index and vector data in host DRAM. As RAG databases grow beyond the available DRAM capacity, keeping the entire database resident becomes impractical. To support these larger datasets, Milvus utilizes memory mapping to retain only lightweight metadata in DRAM while accessing index and vector pages from SSD-backed files~\cite{wang2021milvus,guo2022manu}. A page enters the page cache only when an ANN traversal accesses it. If the page is resident, the traversal continues from memory. Otherwise, a page fault triggers a storage-backed page-in. 

Concurrent sub-searches access many pages irregularly, competing for limited page-cache capacity and storage I/O bandwidth. Repeated eviction and refetching cause page-cache thrashing, while simultaneous page-ins contend for the storage path, adding latency beyond graph traversal.
\subsection{Challenges and Design Implications}
\begin{figure}[t]
\centering
\includegraphics[width=0.95\columnwidth]{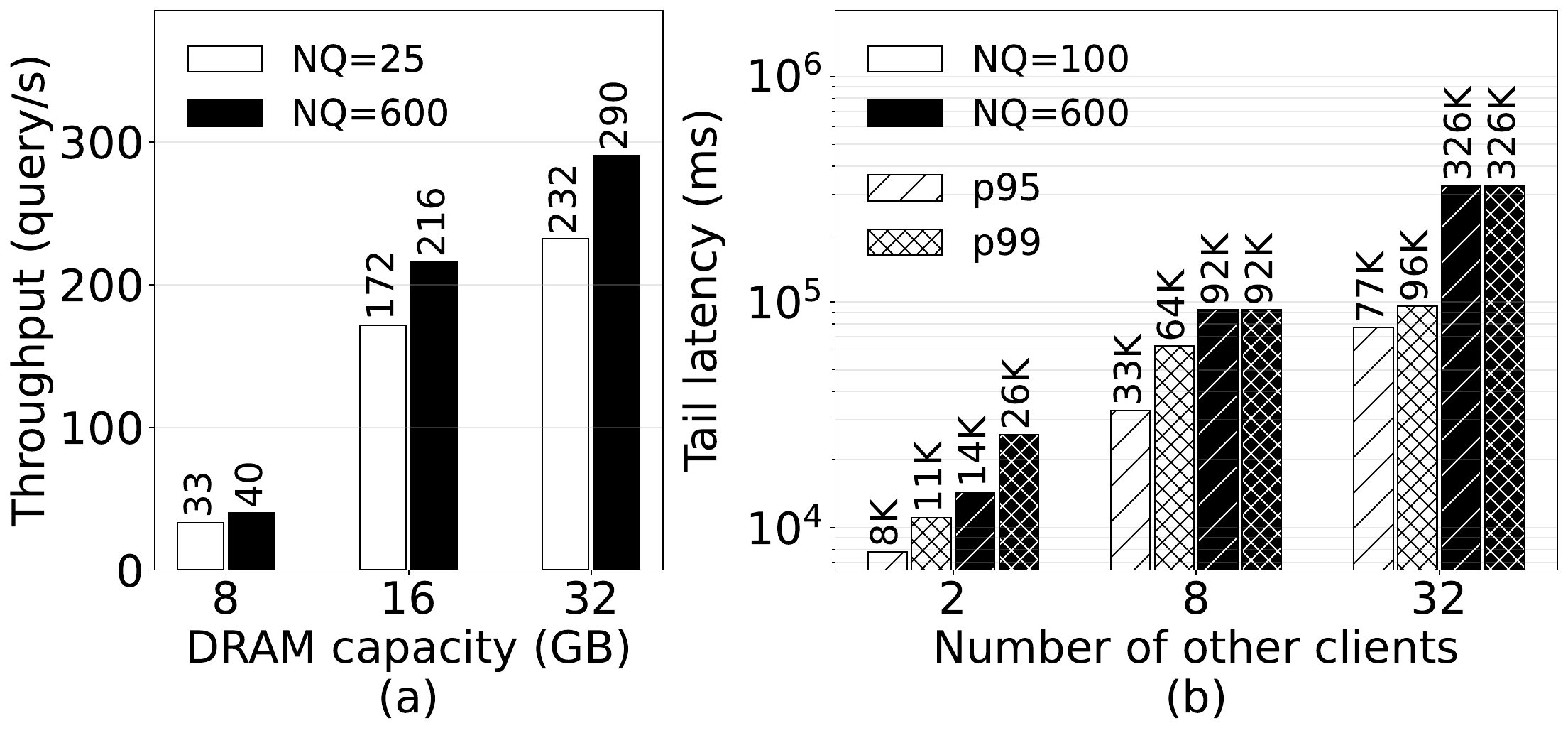}
\caption{Performance of HNSW serving a 64-GB database with limited host DRAM. (a) Throughput across DRAM capacities and query batch sizes. (b) Batch-25 target-client p95/p99 latency with increasing background clients.}
\label{fig:dram_sensitivity}
\end{figure}

\para{Motivating Observation} Figure~\ref{fig:dram_sensitivity}(a) shows a HNSW index on SIFT100M across DRAM capacities and query batch sizes $n_q$. More DRAM enlarges the page-cache set, whereas larger batches expose query-level parallelism and amortize per-batch overhead. Consequently, larger batches achieve higher QPS, motivating a bulk-oriented execution path. 

Figure~\ref{fig:dram_sensitivity}(b) holds a batch-25 client as the latency-sensitive target and varies the number of other clients. The other clients issue batch-100 or batch-600 requests. As the number of other clients increases, the target client's p95 and p99 latency rise substantially, showing how throughput-oriented requests degrade small-batch performance through shared host resources. Overall QPS benefits from larger batches, but unrestricted batch growth can substantially degrade small-batch latency under shared-client workloads. This trade-off motivates an execution design that scales bulk throughput while isolating latency-sensitive small requests.

\para{Design Requirements} The observations lead to three design requirements. Fine-grained demand paging must be replaced with bulk movement at the segment granularity. Because large batches prioritize aggregate throughput over per-query latency, they should be executed on a highly parallel, many-core substrate rather than one optimized primarily for host-CPU-class single-core performance. Latency-sensitive small batches, in contrast, require a separate CPU execution path. Both paths must compute directly on a shared data representation without copying search data between host and device memories, thereby avoiding redundant memory consumption.

\begin{figure}[htbp]
\centering
\SmartIncludeGraphics{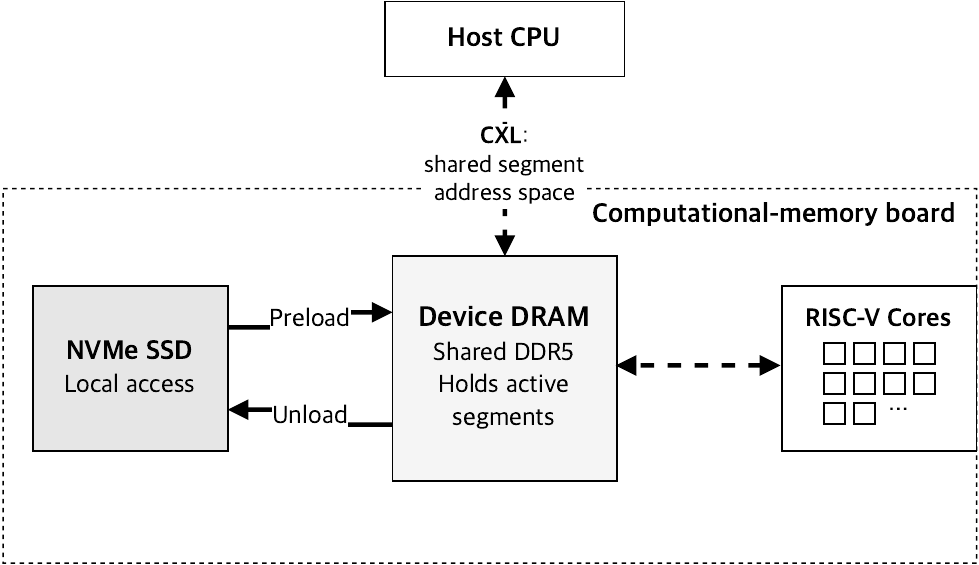}{0.95\columnwidth}{2.0in}
\caption{\cmsflash used as the execution substrate for \name. Multiple RISC-V cores access device DRAM, which is exposed to the host through \cxl.}
\label{fig:xcena_board}
\end{figure}

\para{Target Device} Figure~\ref{fig:xcena_board} shows the XCENA MX1P device used as our \cmsflash: a \cxl~\cite{cxlspec2023} Type-3 device that combines flash-backed capacity, device DRAM, and many RISC-V cores. Its many-core design benefits bulk batches that expose parallelism across queries and segments, whereas its modest per-core performance makes the device path less suitable for small requests. Device-side bulk search remains close to active-segment data, while \cxl exposes device DRAM to the host CPU, allowing both paths to access the same data representation.
  
\para{Our Design} Guided by these requirements and enabled by the target device's shared memory and many-core resources, \name maps its design onto four cooperating roles. \segmentmanager tracks each active segment. An active segment contains the ANN index and vector payload staged in device DRAM. \segmentmanager keeps the segments required by large batches resident ahead of time. Once a segment is resident, \cmsflash runs its sub-searches for each query and segment on the device cores, while \router sends latency-sensitive small batches to the host CPU. This CPU path needs no second copy of the data because \memflash exposes the same CXL-visible segment to both sides. \segmentmanager also holds back device-DRAM slots and preload bandwidth for the CPU path after the bulk look-ahead window is satisfied.

%% file: contents/3-design.tex
\section{\name Design}

\subsection{Overview}
\name targets memory-constrained vector-search serving in which a dataset exceeds host-DRAM capacity and throughput-oriented large batches coexist with latency-sensitive small batches. \name decouples vector-search capacity from host DRAM by storing ANN indexes and vectors in flash and explicitly staging active segments in device DRAM.

\begin{figure}[t]
\centering
\includegraphics[width=0.95\linewidth]{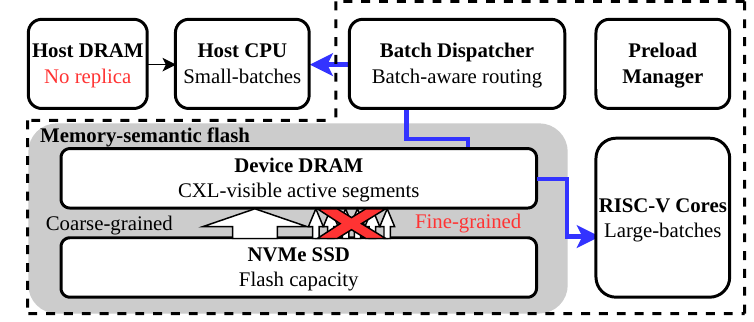}
\caption{\name overview}
\label{fig:architecture_overview}
\end{figure}

As Figure~\ref{fig:architecture_overview} shows, \memflash exposes this flash-backed segment space to the host CPU over \cxl, and device DRAM holds the active segments so that both the host CPU and device-side RISC-V cores work from the same memory; host DRAM does not need a duplicate index or vector store. \router prioritizes device-side execution and uses RISC-V cores utilization to determine when to route the smallest pending batch to the host CPU, while \segmentmanager keeps the bulk-critical active segments required by the device path resident. Both paths return only per-segment top-$k$ results, which the host merges.

\subsection{\cmsflash for Bulk Search}

\para{Segment-Oriented Data Movement} 
Milvus \mmapio relies on demand paging: the host operating system places an index or vector page in DRAM only after ANN traversal accesses the page. When database capacity grows without a commensurate increase in host DRAM, the page cache retains a smaller fraction of the working set and demand paging generates greater host-memory contention. Milvus also fans each query batch out into segment-level sub-searches. A larger segment count can increase the number of concurrent sub-searches, further amplifying page-cache and TLB contention under irregular ANN access. \name instead transfers the ANN index and vector payload required for a search from flash to device DRAM in bulk at segment granularity. Whole-segment transfers are therefore better suited to high-fan-out search: explicit, device-managed data movement replaces demand-paged fine-grained accesses during ANN traversal. As a result, even when the database greatly exceeds host DRAM capacity, this execution path avoids host-OS page-cache management and page-fault handling.

\para{Computation Offloading} Once a segment is resident in device DRAM, \name offloads its query--segment sub-searches to the \cmsflash. A large batch provides many independent query--segment pairs, which are distributed across the device's RISC-V cores. Each core performs ANN traversal or probing, distance computation, and local top-$k$ selection near the resident index and vectors. The host dispatches work and merges per-segment top-$k$ lists, while the device returns only these compact results rather than vector data, index pages, or complete candidate payloads. Device-side parallel execution increases bulk-search QPS while avoiding transfers of index pages, vectors, and full candidate payloads to the host.

\subsection{Batch-Aware Dual-Path Execution}

\para{\memflash: Shared Access to Active Segments} \memflash provides a single data representation shared by the host-CPU and device-side execution paths. Flash provides its capacity, while device DRAM serves as the working-memory tier for active data. The host CPU accesses each resident segment through a \cxl mapping, while the device-side execution path accesses it locally. Consequently, routing work between the two paths requires neither a host-DRAM replica nor copying or reformatting the ANN index and vector payload.

\begin{figure}[t]
\centering
\SmartIncludeGraphics{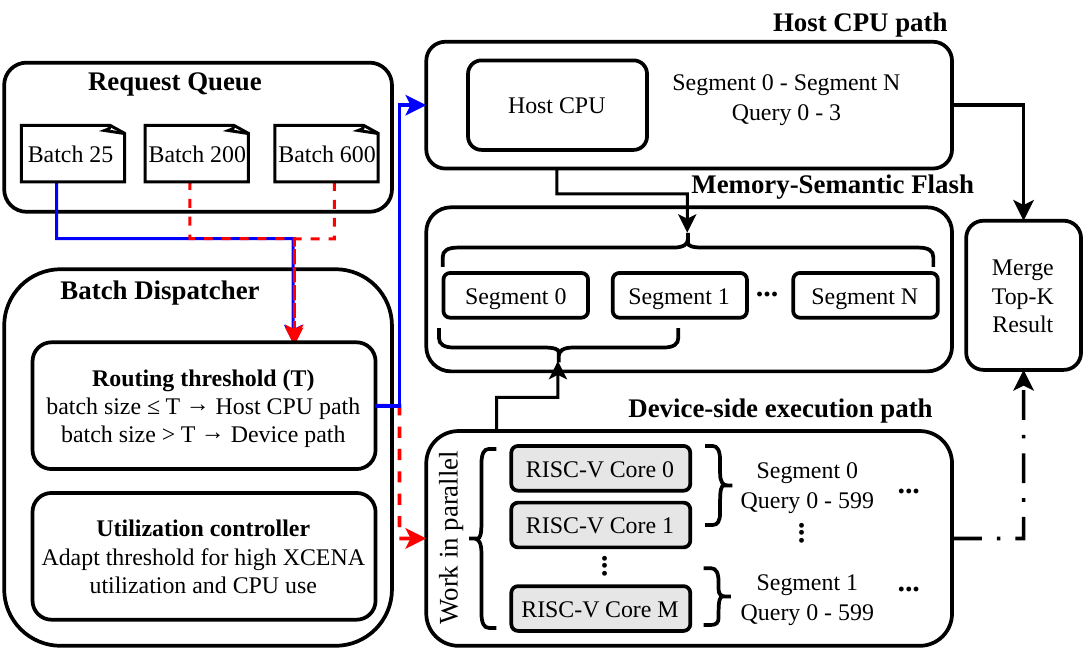}{0.95\linewidth}{2.2in}
\caption{Batch-aware dual-path routing in \router.}
\label{fig:dual_path_execution}
\end{figure}

\para{\router: Batch-Aware Routing} Large batches generate enough query--segment sub-searches to exploit device-side parallelism, whereas latency-sensitive small batches contribute relatively few sub-searches to each target segment and can require a wider, changing set of segments. As Figure~\ref{fig:dual_path_execution} illustrates, \router directs the latter requests to the host CPU while prioritizing the device-side path for bulk work. Because \memflash exposes the same resident segments to both paths, this decision changes the execution location without copying the index or vector payload. \router maintains a batch-size routing threshold based on device-side RISC-V core utilization. When this utilization exceeds 80\% and the host CPU path is idle, the dispatcher adjusts the threshold on each new request arrival so that the smallest pending batch is routed to the host CPU. This adjustment diverts only the lowest-batch work and keeps the remaining requests on the device-side path. Both paths return only per-segment top-$k$ results for host-side merging.

\begin{figure}[t]
\centering
\includegraphics[width=0.95\linewidth]{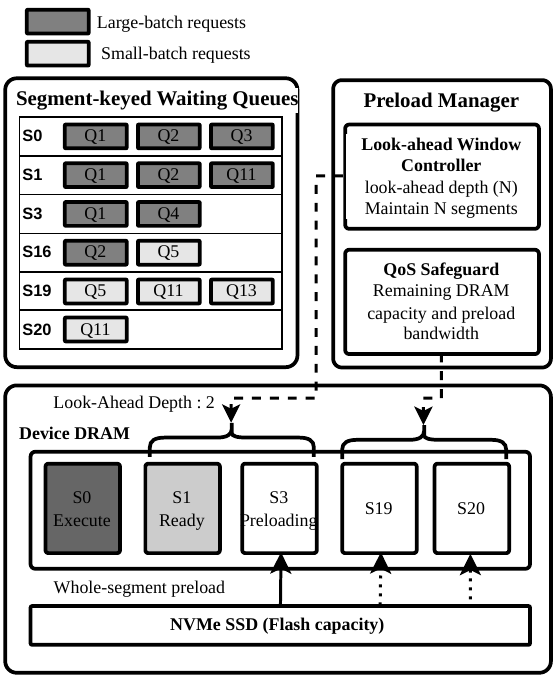}
\caption{Illustration of \segmentmanager operation.}
\label{fig:segment_preload}
\end{figure}

\para{\segmentmanager: Explicit Segment Residency} 
The \cmsflash device executes searches from device DRAM. It therefore brings a segment's ANN index and vector payload into device DRAM before computation; we refer to this whole-segment transfer as preloading. \name uses a device-defined $1$\,GB segment as the unit of preloading, residency, placement, and search, and \segmentmanager maintains segment-keyed waiting queues for pending sub-searches. A large batch fans out many sub-searches to each bulk-critical unit. Consequently, a small resident set can provide enough parallel work to keep the device-side RISC-V cores computation-bound.

The look-ahead window controller adjusts its depth $N$ toward the smallest value that sustains computation-bound execution. For every preloaded segment, it measures the interval from becoming resident to exhausting its queued large-batch sub-searches, and compares this consumption time with the flash-to-device-DRAM refill latency. If a segment is consumed before its replacement is ready, the controller increases $N$; if replacement segments are ready before they are needed, it decreases $N$. The controller therefore maintains only the minimum number of bulk-critical active segments needed to prevent device-side starvation, rather than partitioning device DRAM or retaining a broad fraction of the flash-resident database.

After the look-ahead window has secured $N$ bulk-critical active segments, the QoS safeguard assigns all remaining device-DRAM capacity and flash-to-device-DRAM bandwidth to the host CPU path. Small batches can therefore use the remaining resources to access a wider and changing set of segments, while the device-side execution path retains only the few segments needed to avoid device-side starvation. Resident segments remain shared through \memflash, so a request on either path can use an already resident segment without creating a replica. Before execution, each resident segment contains both the ANN index and vector payload, and the host merges the resulting per-segment top-$k$ lists.

%% file: contents/5-evaluation.tex
\section{Evaluation}
\label{sec:eval}
\subsection{Experimental Setup}

\para{System Configuration} We evaluate \name on a 72-core Intel Xeon 6 P-core (Granite Rapids-SP) processor with 373\,GiB of DDR5 DRAM. The host connects to an XCENA MX1P device~\cite{xcena} configured as a \cmsflash device and uses a Samsung 16\,TB PM1753 NVMe SSD as the backing storage device.

\para{Systems Compared} We configure vanilla Milvus v2.6.9 to use 8 host CPU cores, reflecting a shared CPU allocation rather than exclusive access to the full 72-core socket. Vanilla Milvus serves as our baseline and executes ANN search on the host CPU. We enforce the 8-core allocation using \texttt{systemd-run} scopes.

\begin{table}[t]
\centering
\scriptsize
\resizebox{\columnwidth}{!}{%
\begin{tabular}{@{}lrrrr@{}}
\hline
Dataset & Base Vector & Queries & Dim. & Metric \\\hline
SIFT & 1\,B & 100\,M & 128 & L2 \\
Deep & 1\,B & 1\,M & 96 & L2 \\
Text2Image & 1\,B & 100\,K & 200 & IP \\\hline
\end{tabular}
}
\caption{Datasets and query sets. All vectors are stored as Float32 in Milvus; SIFT vectors are converted from Uint8}  
\label{tab:workloads}
\end{table}

For billion-scale workloads, we limit the baseline to 64\,GiB of host DRAM; for the SIFT100M mixed-workload experiment, we limit it to 16\,GiB. These limits keep the host-memory working set smaller than the dataset and expose the demand-paging and host-memory contention regime targeted by this work. In contrast, \name uses 16\,GiB of device DRAM as the working-memory tier for active segments rather than relying on host DRAM for ANN index and vector residency. \name-Stage explicitly stages whole segments in device DRAM while retaining host-CPU ANN execution. \name uses the utilization-aware \router to adapt its routing threshold, maintaining high XCENA utilization while making use of the host CPU path.

\para{Datasets and Query Sets} We use the billion-scale SIFT1B, Deep1B, Text2Image1B workloads to evaluate search throughput, and 100M-scale datasets for mixed-workload latency experiments. Table~\ref{tab:workloads} summarizes the billion-scale workload configurations. SIFT vectors are converted from their original uint8 representation to Float32 to maintain a consistent vector representation across workloads and match the format used by Milvus. For sustained high-concurrency experiments, we use the 100M-query learn set of SIFT1B, the 1M private query set of Deep1B, and the 100K public query set of T2I1B. In Milvus, HNSW indexes are built with $M=16$ and $\mathit{efConstruction}=200$. All searches use top-$k=10$ and $\mathit{ef}=64$.

\para{Workloads} For search-throughput experiments, 8 clients concurrently issue batch-600 requests. For mixed workloads, one client continuously issues batch-25 requests while 1, 2, 4, or 8 clients issue batch-600 requests. Each client uses an independent Milvus connection and operates in a closed loop. Before each run, we warm up the system using the corresponding query set and batch size. At the end of the timed interval, we stop admitting new requests and report only those completed before the cutoff.

\begin{figure}[t]
\centering
\includegraphics[width=0.95\columnwidth]{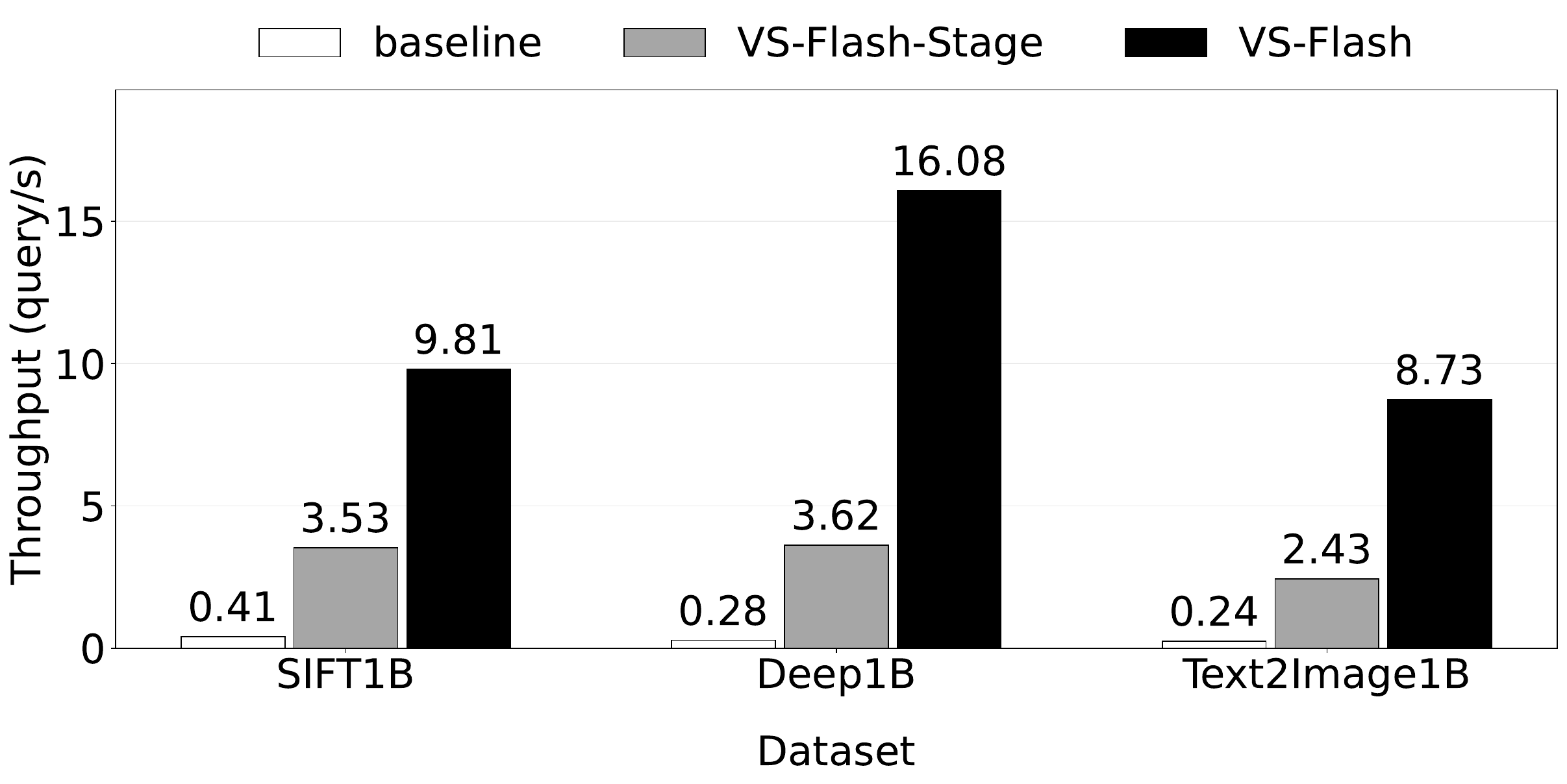}
\caption{Aggregate bulk-search throughput with 64\,GiB of host DRAM. 8 clients concurrently issue batch-600 requests.}
\label{fig:qps}
\end{figure}

\subsection{Performance Results}
\para{Throughput on Billion-Scale Datasets} Figure~\ref{fig:qps} compares vanilla Milvus, \name-Stage, and \name across the three billion-scale datasets. \name improves throughput over vanilla Milvus by 23.9$\times$, 57.4$\times$, and 36.4$\times$ on SIFT1B, Deep1B, and Text2Image1B, respectively, for a 36.8$\times$ geometric-mean improvement. Across the workloads, explicit whole-segment staging provides a 10.4$\times$ average improvement, and device-side execution provides a further 3.5$\times$ average improvement over \name-Stage.

\name achieves high throughput on SIFT1B, Deep1B, and T2I, with the largest gain from device-side RISC-V many-core offloading on Deep1B. Deep1B uses lower-dimensional vectors than the other datasets, shifting the dominant cost from compute-intensive distance evaluation toward the lighter-weight operations of HNSW graph traversal. As a result, the lower per-core compute performance of the device-side RISC-V cores has a smaller effect on search latency, while large batches still expose ample query--segment parallelism for the many-core device.

\begin{figure}[t]
\centering
\includegraphics[width=0.95\columnwidth]{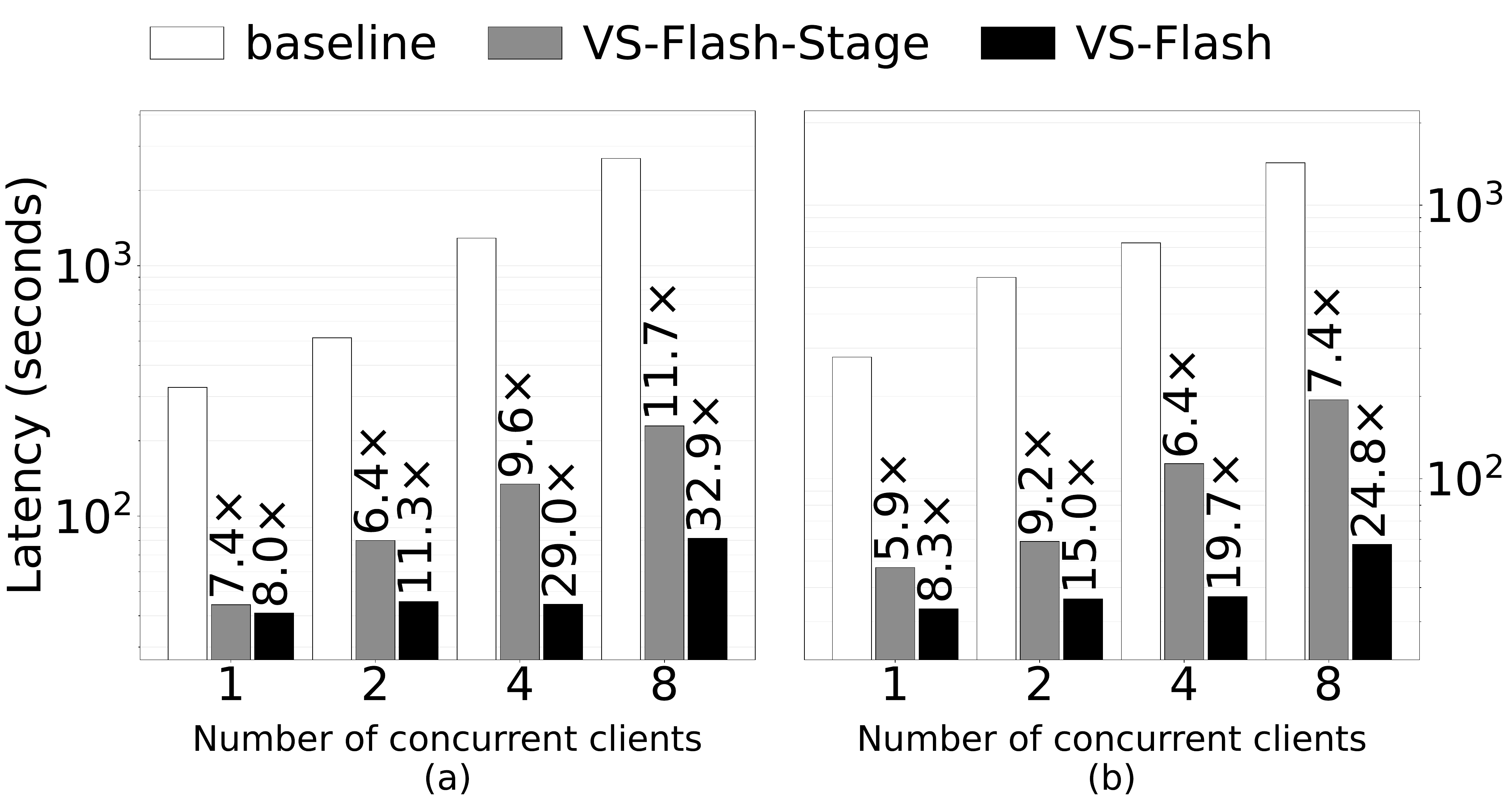}
\caption{Mixed-workload p99 latency on SIFT100M as bulk-client concurrency increases: (a) batch-600 requests and (b) batch-25 requests.}
\label{fig:tail_latency}
\end{figure}

\para{Mixed-Workload Latency} Figure~\ref{fig:tail_latency} reports p99 latency as the number of concurrent batch-600 clients increases. \name-Stage substantially reduces the p99 latency of both batch classes, showing that explicit segment staging avoids the demand-paging and storage contention incurred by vanilla Milvus. \name provides the largest reduction under high bulk pressure: with eight batch-600 clients, it reduces batch-600 and batch-25 p99 latency by 32.9$\times$ and 24.8$\times$, respectively. The benefit grows with bulk concurrency because \router directs latency-sensitive batch-25 requests to the host CPU rather than allowing them to queue behind throughput-oriented device-side work. This separation also protects the bulk path: by removing small requests from the device-side queue, the RISC-V cores remain available for batch-600 execution. Thus, \name not only eliminates the demand-paging bottleneck, but also sustains high-throughput bulk search while protecting latency-sensitive requests under severe mixed-workload interference.
  
\begin{figure}[t]
\centering
\includegraphics[width=0.95\columnwidth]{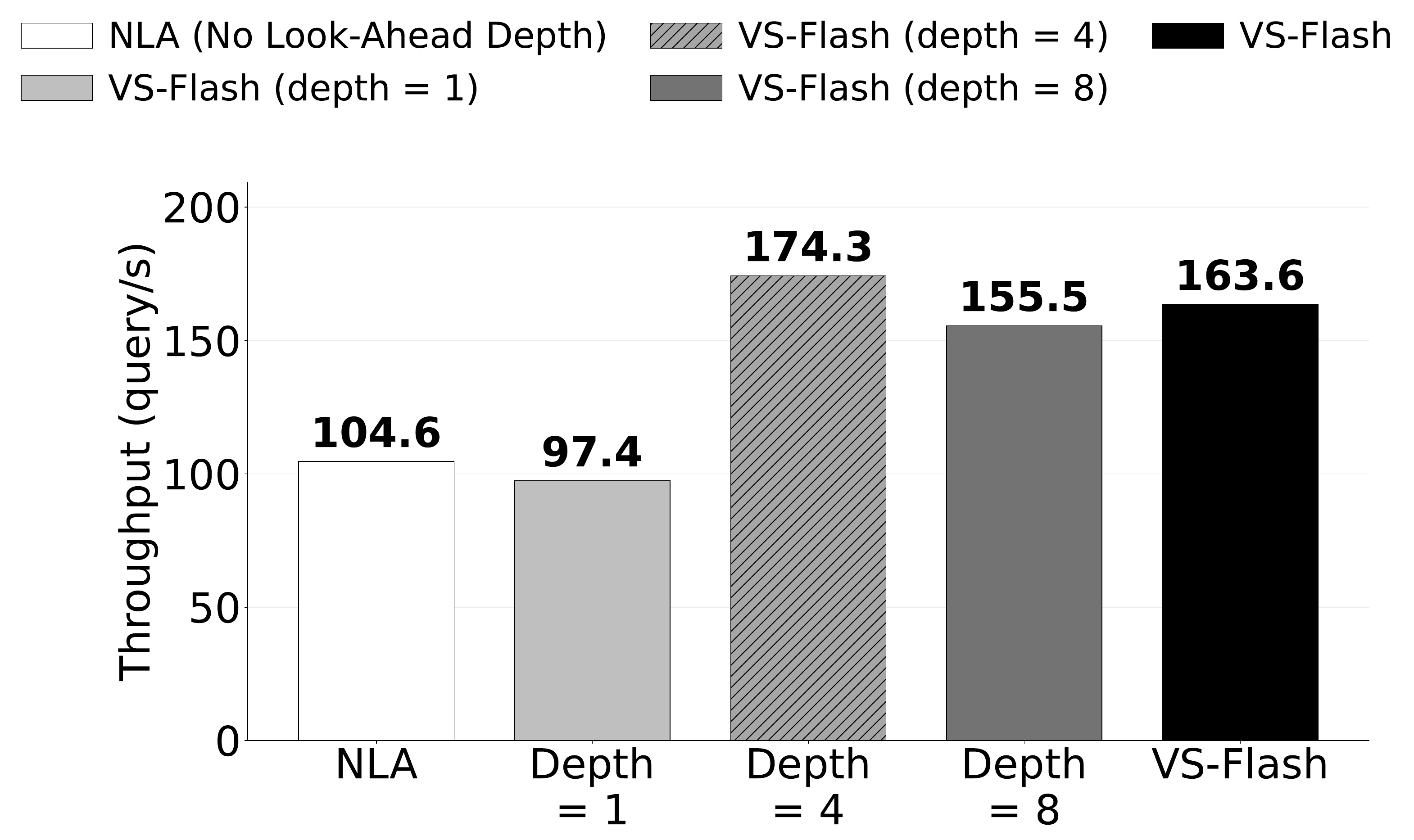}
\caption{Effect of the look-ahead policy under the mixed workload.}
\label{fig:lookahead_sensitivity}
\end{figure}

\para{Preload-Management Sensitivity} Figure~\ref{fig:lookahead_sensitivity} evaluates \name's look-ahead window controller against no look-ahead (NLA) and several fixed window depths. NLA does not manage look-ahead preloading, causing the \cmsflash path and the CPU path to contend for device-DRAM capacity and flash-to-device-DRAM bandwidth.

A fixed depth of $N=1$ underperforms NLA because reserving only a single look-ahead segment is insufficient to keep pace with \cmsflash consumption, causing device-side starvation when the next segment is not ready, yet the preloaded segment can still be used by the host CPU path. QPS peaks at $N=4$. At $N=8$, additional device-DRAM capacity is reserved for segments that provide no further device-side throughput benefit, while substantially reducing the capacity available to the host CPU path; aggregate throughput therefore decreases. These results show that selecting an appropriate window depth is essential.

\name's look-ahead window controller uses an initial depth of $N=2$ in this experiment and adjusts the depth based on the observed segment consumption rate and p95 preload latency. Although \name's QPS is lower than that of a manually fixed $N=4$ because execution begins from a different depth, the controller converges to $N=4$, the best fixed-depth setting for this workload. \name achieves 56.4\% higher throughput than NLA and within 6.1\% of the manually selected configuration, demonstrating that \name identifies an effective preload allocation without offline tuning.
  
%% file: contents/7-relatedwork.tex
\section{Related Work}
\label{sec:rel}

\para{ANN Indexing Algorithms} ANN indexes such as HNSW, IVF, DiskANN, FreshDiskANN, and SPANN optimize the accuracy--speed trade-off through graph traversal, partition probing, and disk-resident structures~\cite{hnsw, faiss, DiskAnn, freshdiskann, spann, turbocharching}. Previous works optimize index efficiency, whereas \name retains existing ANN indexes and their configured search parameters, improving serving performance without sacrificing recall through index or algorithm changes.

\para{Vector Database Systems} Vector database systems such as Milvus, Manu, and VBASE support indexing, storage, query processing, and distributed scaling~\cite{wang2021milvus, guo2022manu, VBASE, surveyofVDBMS, surveyTech}. Although they accommodate large datasets, they do not explicitly address page-residency and contention problems in MMap-based serving. \name extends Milvus with a role-separated execution model for this setting.

\para{Tail Latency and Multi-Tenant Interference} Prior work shows that contention for shared CPU, memory, and I/O resources can severely degrade individual tenants' tail latency~\cite{tailatscale, Pisces, sqlvm, delta, curator, Tales}. Existing approaches primarily consider general purpose storage or relational workloads, rather than vector-search interference caused by irregular ANN access, search fan-out, and page faults. \name addresses this gap for batch-heterogeneous vector-search workloads.

\para{Near-Data and CXL-Based Vector Search} Prior work moves ANN computation closer to data to reduce data movement and resource pressure~\cite{compstorage, recssd, smartrec, ndsearch, kvcsd}. ANSMET accelerates distance computation with DIMM-based near-memory processing and hybrid early termination~\cite{ansmet2025}. Cosmos offloads full ANN search to CXL-embedded cores, albeit evaluated in simulation, while FusionANNS partitions SSD-resident ANNS between CPU filtering and GPU re-ranking~\cite{cosmos2025, fusionanns2025}. 

Other systems explore CXL-memory offloading, hardware acceleration, CPU/GPU/NPU partitioning, or SSD I/O pipelining~\cite{udon2024, axle2026, falcon2025, reflux2026, vectorliterag2025, ascendrabitq2026, pipeann2025}. Unlike these systems, \name dynamically routes requests by batch size between the host CPU and device path to protect tail latency.

%% file: contents/6-conclusion.tex
\section{Conclusion}
As vector-search databases outgrow host DRAM, MMap-based serving relies on demand paging during irregular ANN traversal. This approach creates host-memory and storage contention, particularly when throughput-oriented large batches coexist with latency-sensitive small batches. \name replaces demand paging with explicit whole-segment residency in device DRAM. It executes bulk searches on device-side cores while allowing the host CPU to access the same resident segments for small requests without a host-DRAM replica. Our evaluation shows that this organization improves bulk-search throughput and protects small-batch performance under mixed workloads.